# Covalent Hybrid of Spinel Manganese-Cobalt Oxide and Graphene as Advanced Oxygen Reduction Electrocatalysts

Yongye Liang,[1,§] Hailiang Wang,[1,§] Jigang Zhou,[2] Yanguang Li,[1] Jian Wang,[2] Tom Regier[2] and Hongjie Dai[*,1]

[1]Department of Chemistry, Stanford University, Stanford, CA 94305, USA.
[2]Canadian Light Source Inc., Saskatoon, SK S7N 0X4, Canada.

**ABSTRACT:** Through direct nanoparticle nucleation and growth on nitrogen doped, reduced graphene oxide sheets and cation substitution of spinel $Co_3O_4$ nanoparticles, a manganese-cobalt spinel $MnCo_2O_4$/graphene hybrid was developed as a highly efficient electrocatalyst for oxygen reduction reaction (ORR) in alkaline conditions. Electrochemical and X-ray near edge structure (XANES) investigations revealed that the nucleation and growth method for forming inorganic-nanocarbon hybrid results in covalent coupling between spinel oxide nanoparticles and N-doped reduced graphene oxide (N-rmGO) sheets. Carbon K-edge and nitrogen K-edge XANES showed strongly perturbed C-O and C-N bonding in the N-rmGO sheet, suggesting the formation of C-O-metal and C-N-metal bonds between N-doped graphene oxide and spinel oxide nanoparticles. Co L-edge and Mn L-edge XANES suggested substitution of $Co^{3+}$ sites by $Mn^{3+}$, which increased the activity of the catalytic sites in the hybrid materials, further boosting the ORR activity compared to the pure cobalt oxide hybrid. The covalently bonded hybrid afforded much greater activity and durability than the physical mixture of nanoparticles and carbon materials including N-rmGO. At the same mass loading, the $MnCo_2O_4$/N-graphene hybrid can outperform Pt/C in ORR current density at medium overpotentials with superior stability to Pt/C in alkaline solutions.

## INTRODUCTION

The oxygen reduction reaction (ORR) has been one of the focuses of electrochemistry in the past decades owing to its importance and continued challenges to a variety of energy applications including fuel cells[1] and metal-air batteries[2]. Platinum based materials are known to be the most active catalysts for ORR in both acidic and alkaline conditions[3], but come at a price of high cost and limited stability[3c]. Compared to its behaviors in acidic conditions, ORR in alkaline conditions is more facile and can be catalyzed by a boarder range of materials more stable in base than in acid[4]. Metal oxides[5], nitrogen coordinated metal on carbon matrices[6] and doped carbon materials[7] have been investigated as ORR catalysts in alkaline solutions. However, it remains a challenge to develop an ORR catalyst with a high activity comparable to Pt, but with higher durability and much lower cost.

Mixed valence oxides of transition metals with a spinel structure are an important class of metal oxides that exhibit ORR catalytic activity in alkaline conditions[8]. Due to the activity, low cost, simple preparation and high stability, cobalt based spinel oxides have been investigated extensively as electrocatalysts[9]. In particular, substituted $Co_3O_4$ with Ni, Cu and Mn have shown high activity and stability as electrocatalysts for ORR[9,10]. However, despite these efforts, substituted spinel oxides still exhibit a much lower mass activity when compared to Pt based materials. For example, in 6 M KOH at ~0.2 V vs. Hg/HgO, a $MnCo_2O_4$-carbon black catalyst at 14 mg/cm$^2$ loading gave current density of 300 mA/cm$^2$ at 60 ℃, while Pt/CNT/C at 0.1 mg/cm$^2$ loading gave current density of 125 mA/cm$^2$ at 25 ℃.[11] Despite the low cost, there is a maximum acceptable electrode loading of the catalyst due to the resistivity in oxygen and electrolyte transport[11]. Thus, it is highly desirable to develop spinel oxide catalysts with much improved mass/volume catalytic activity.

Recently, we showed that the direct nucleation, growth and anchoring of nanocrystals of various shapes on graphene oxide (GO) sheets result in a series of hybrid materials with optimal electrical and chemical coupling between the nanoparticles and the GO substrates[12]. In particular, we found that direct nucleation and growth of $Co_3O_4$ nanoparticles on N-doped (with carbon atoms covalently bonded to nitrogen containing functional groups), reduced graphene oxide (N-rmGO) afforded intimate bonding and synergetic coupling effects, leading to significantly higher electrocatalytic ORR activity than either $Co_3O_4$ or N-doped graphene alone or their physical mixture[13]. The $Co_3O_4$/graphene hybrid catalyst was superior to Pt in terms of stability and durability, but lower in ORR activity than Pt in basic solutions[13]. Here, we report the synthesis of a $MnCo_2O_4$/graphene hybrid for high performance ORR electrocatalysis, by taking advantage of the higher activity of $MnCo_2O_4$ than pure $Co_3O_4$ and the strong coupling with N-doped graphene. Spinel manganese cobalt oxide could be synthesized by various preparation techniques, such as ceramic method[9], organic co-precipitation[14]. Recently, a rapid reduction-recrystallization method was developed to form nanocrystalline spinel oxide from amorphous MnO$_2$ at room temperature[15]. In order to achieve selective growth of MnCo$_2$O$_4$ nanoparticles on graphene sheets, we used a simple two-step solvothermal method to synthesize MnCo$_2$O$_4$/graphene hybrid material. The hybrid showed more positive onset and peak potential for ORR and a greater electron transfer number than the corresponding physical mixture of MnCo$_2$O$_4$ nanoparticles and N-doped graphene sheets. Not only did the MnCo$_2$O$_4$/graphene exhibit higher activity than the Co$_3$O$_4$/graphene hybrid, it also outperformed Pt/C in terms of the ORR current density for the same mass loading in alkaline solutions under medium to high overpotentials. This was accompanied by superior stability to Pt/C. X-ray spectroscopy revealed the covalent coupling of the MnCo$_2$O$_4$ nanoparticles with N-doped graphene, suggesting that strongly coupled hybrid materials offer a promising strategy for advanced electrocatalysts.

## RESULTS AND DISCUSSION

A two step method was developed to synthesize MnCo$_2$O$_4$ nanoparticles on graphene oxide sheets[13]. In the first step of the reaction, Co(OAc)$_2$ and Mn(OAc)$_2$ at certain ratio were reacted with mildly oxidized graphene oxide (mGO) in an ethanol/water NH$_4$OH solution at 80 °C (see Experimental Section for details) to selectively form a uniform coating of hydrolyzed precursors on mGO sheets without free growth in solution. This step was referred to as the nucleation step, during which ammonium hydroxide was added to mediate the nucleation of metal species onto the functional groups of mGO and provide a source of nitrogen doping of mGO[13]. Subsequently, hydrothermal treatment at 150 °C was performed to afford N-doped, reduced mGO (N-rmGO) and crystallization of the metal oxide nanoparticles on N-rmGO, yielding the designed spinel manganese-cobalt oxide/N--graphene hybrid material. We aimed at making MnCo$_2$O$_4$/N-rmGO with Mn/Co = 1/2 by controlling the reactant ratio of Co(OAc)$_2$/Mn(OAc)$_2$, since such a composition had been shown to be more active than other Mn substituted cobalt oxide[16].

Scanning electron microscopy (SEM) and transmission electron microscopy (TEM) clearly revealed the formation of nanocrystals (average size of ~5 nm) on graphene sheets (Figure 1a&b). X-ray diffraction (XRD) showed that the synthesized nanocrystals were in cubic spinel phase and the peaks were slightly shifted to smaller 2θ angles compared to pure Co$_3$O$_4$ due to substitution of larger size Mn cations in a solid solution (Figure 1c). The lattice fringes of the nanocrystals revealed by high resolution TEM were consistent with the MnCo$_2$O$_4$ crystal structure (Figure 1d). Both X--ray photoelectron spectroscopy (XPS) (Figure 1e) and energy dispersive spectroscopy (EDS) (Figure. 1f) showed that the Co/Mn ratio was ~2 in the hybrid, close to the designed ratio. There was ~ 4 at% nitrogen in the MnCo$_2$O$_4$/N-rmGO hybrid, with N incorporated in reduced graphene oxide[13] instead of in MnCo$_2$O$_4$ (Figure 1e and high resolution N1s spectrum in Fig. S1). The amount of N-rmGO in the final hybrid was ~20% by mass measured by thermal analysis.

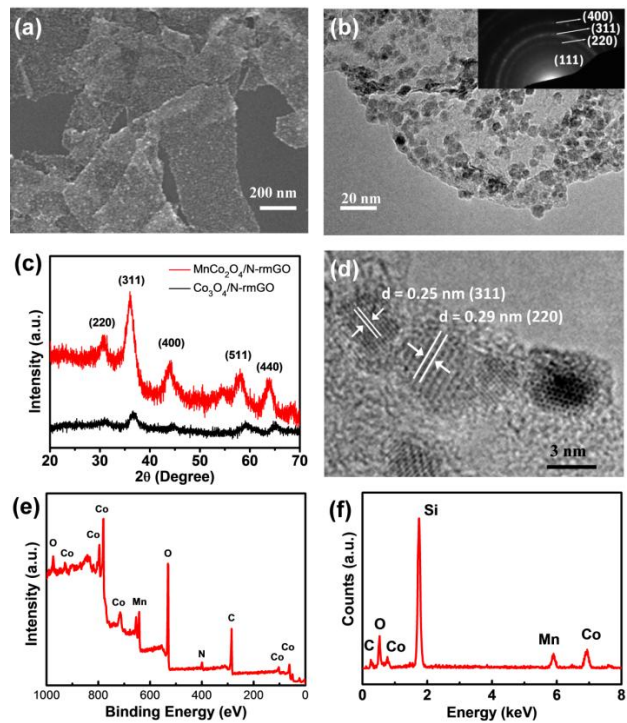

**Figure 1.** (a) Low magnification SEM image and (b) Low magnification TEM image with an inset of electron diffraction pattern of MnCo$_2$O$_4$/N-rmGO hybrid deposited on a silicon substrate and holey carbon grid respectively. (c) XRD spectra of compacted films of MnCo$_2$O$_4$/N-rmGO hybrid (red) and Co$_3$O$_4$/N-rmGO hybrid (black). (d) A high magnification TEM image of MnCo$_2$O$_4$/N-rmGO hybrid. (e) An XPS spectrum of MnCo$_2$O$_4$/N-rmGO hybrid. (f) An EDS spectrum of MnCo$_2$O$_4$/N-rmGO hybrid. The molar ratio of Co/Mn is about 2.0-2.2, which is consistent to the reagent ratio.

The electrocatalytic activity of MnCo$_2$O$_4$/N-rmGO hybrid for ORR was first characterized by cyclic voltammetry (CV) in 1 M KOH on a glassy carbon electrode and compared with several other catalysts (Figure 2a). The ORR onset potential and peak potential of MnCo$_2$O$_4$/N-rmGO hybrid were 0.95 V and 0.88 V respectively vs. the reversible hydrogen electrode (RHE), ~ 20 mV more positive than those of pure Co$_3$O$_4$/N-mGO hybrid (onset potential of 0.93 V and peak potential of 0.86 V vs. RHE). Note that the ORR peak potential for Pt/C catalyst (20 wt% Pt on Vulcan XC-72) was located at 0.90 V, only ~20 mV more positive than that of MnCo$_2$O$_4$/N-rmGO hybrid (Fig. S2a). This suggested that Mn substitution in the cobalt oxide/graphene hybrid enhanced the ORR catalytic activity to approach that of Pt/C in 1 M KOH. Free MnCo$_2$O$_4$ nanoparticles alone showed very poor ORR catalytic activity (Fig. S2b). N-rmGO sheets alone without any metal oxides showed certain ORR catalytic activity, but the ORR onset potential (0.89 V) and peak potential (0.82 V) were more negative than the hybrid materials (Figure 2a). The physical mixture of MnCo$_2$O$_4$ nanoparticles and N-rmGO showed improved catalytic activity compared to each component alone, but was still inferior to the hybrid in activity with an onset potential of 0.91 V and peak potential of 0.84 V. This suggested that similar to the Co$_3$O$_4$/N-rmGO case[13], the high catalytic activity of MnCo$_2$O$_4$/N-rmGO is facilitated by the strong coupling between MnCo$_2$O$_4$ nanoparticles and N-rmGO. It should be pointed out that the physical mixture of free $MnCo_2O_4$ nanoparticles and Vulcan carbon black ($MnCo_2O_4$ + CB), a conventional way to prepare oxide electrode for ORR[5b, 10], showed much lower activity with more negative peak potential (0.78 V) than the $MnCo_2O_4$ + N-rmGO mixture and the $MnCo_2O_4$/N-rmGO hybrid.

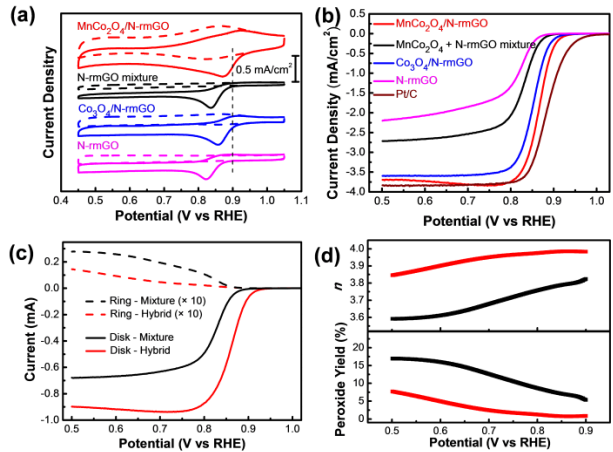

$MnCo_2O_4$/N-rmGO hybrid material[17]. Figure 2c shows the disk and ring currents recorded at 1600 rpm in 1 M KOH for the $MnCo_2O_4$/N-rmGO hybrid and the physical mixture of $MnCo_2O_4$ nanoparticles and N-rmGO sheets. The hybrid sample exhibited a higher disk current ($O_2$ reduction) and a smaller ring current (peroxide oxidation) than the mixture. The percentage of peroxide species with respect to the total oxygen reduction products and the electron reduction number ($n$) calculated from RRDE curves in Figure 2c are shown in Figure 2d. The yield of peroxide species was less than 10% over the measured potential range for the $MnCo_2O_4$/N-rmGO hybrid catalyst, significantly lower than that of ~20% for the mixture. The average electron transfer number was ~3.9 from 0.9 V to 0.5 V for the hybrid catalyst, and was ~3.7 for the mixture. The electron transfer number ($n$) obtained from the Koutecky-Levich plots[18] based on the RDE measurements (Fig. S4) was consistent with RRDE result, indicating that the ORR catalyzed by $MnCo_2O_4$/N-rmGO was mainly through the four electron (4 e) pathway.

The $MnCo_2O_4$/N-rmGO hybrid also showed high ORR catalytic activity in 0.1 M KOH. CV and RDE measurements with $MnCo_2O_4$/N-rmGO in 0.1 M KOH revealed more positive peak potential and half way potential than other catalysts including pure $Co_3O_4$/N-rmGO (Fig. S5 a & b). Both RDE and RRDE measurements suggested the ORR catalyzed by the Mn substituted cobalt oxide hybrid was dominated by a 4-electron reduction pathway in 0.1 M KOH (Fig. S5 b-d).

**Figure 2.** (a) CV curves of $MnCo_2O_4$/N-rmGO hybrid, $MnCo_2O_4$+N-rmGO mixture, $Co_3O_4$/N-rmGO hybrid and N-rmGO on glassy carbon electrodes in $O_2$-saturated (solid line) or $N_2$-saturated (dash line) 1 M KOH. The peak position of Pt/C was shown as a dashed line for comparison. (b) Rotating-disk electrode voltammograms of $MnCo_2O_4$/N-rmGO hybrid, $MnCo_2O_4$+N-rmGO mixture, $Co_3O_4$/N-rmGO hybrid, N-rmGO and Pt/C in $O_2$-saturated 1 M KOH at a sweep rate of 5 mV/s at 1600 rpm. (c) Rotating ring-disk electrode voltammogram of $MnCo_2O_4$/N-rmGO hybrid and $MnCo_2O_4$+N-rmGO physical mixture in $O_2$-saturated 1 M KOH at 1600 rpm. The disk potential was scanned at 5 mV/s and the ring potential was constant at 1.3 V vs. RHE. (d) Percentage of peroxide (bottom) with respect to the total oxygen reduction products and the electron transfer number ($n$) (top) of $MnCo_2O_4$/N-rmGO hybrid and $MnCo_2O_4$+N-rmGO mixture at various potentials based on the corresponding RRDE data in (c). Catalyst loading was 0.10 mg/cm$^2$ for all samples.

Similar trends in the ORR activities were observed in rotating disk electrode (RDE) measurements using the catalysts. The RDE linear sweeping voltammograms in $O_2$ saturated 1 M KOH solution (with the baseline recorded in $N_2$ saturated 1 M KOH subtracted) at a rotation rate of 1600 rpm were compared in Figure 2b and Fig. S3a. The $MnCo_2O_4$/N-rmGO hybrid catalyst outperformed other catalysts including the $MnCo_2O_4$ + N-rmGO mixture, $Co_3O_4$/N-rmGO hybrid and N-rmGO in terms of disk current density and half-wave potential. The half-wave potential difference between the $MnCo_2O_4$/N-rmGO hybrid and the Pt/C catalyst was only 20 mV at the same mass loading. The high catalytic activity of $MnCo_2O_4$/N-rmGO hybrid was also revealed by the small Tafel slope of kinetic current, which was 36 mV/decade in 1M KOH at low overpotential, similar to that of the $Co_3O_4$/N-rmGO hybrid (38 mV/decade) (Figure S3b). The Tafel slope is close to $2.303(2RT/3F)$ V/decade ($R$, universal gas constant; $F$, faraday constant), suggesting that the ORR rate limiting step is related to the protonation of $O_2^-$ on the active sites of catalyst[10].

Rotating ring disk electrode (RRDE) measurements were performed to determine the ORR pathways catalyzed by

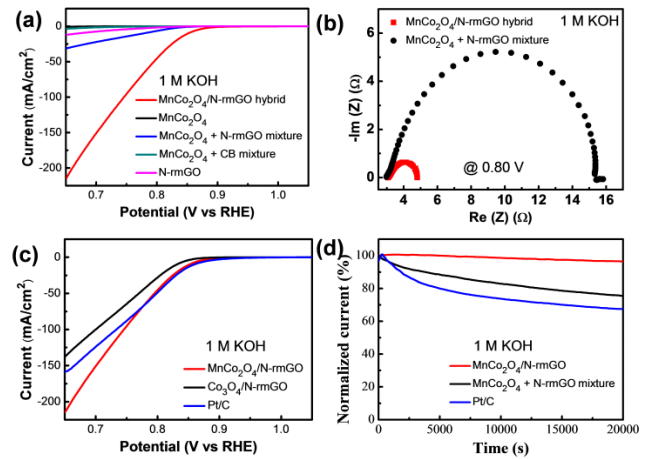

**Figure 3.** (a) Oxygen reduction polarization curves of $MnCo_2O_4$/N-rmGO hybrid, $MnCo_2O_4$, $MnCo_2O_4$ + N-rmGO mixture, $MnCo_2O_4$ + CB mixture and N-rmGO loaded on carbon fiber papers in 1 M KOH electrolyte. (b) Impedance spectra of $MnCo_2O_4$/N-rmGO hybrid and the mixture of $MnCo_2O_4$ and N-rmGO at 0.8 V vs. RHE. (c) Oxygen reduction polarization curves of $MnCo_2O_4$/N-rmGO hybrid, $Co_3O_4$/N-rmGO hybrid, and Pt/C on carbon fiber papers in 1 M KOH electrolyte. (d) Chronoamperometric responses of $MnCo_2O_4$/N-rmGO hybrid, $MnCo_2O_4$ + N-rmGO mixture and Pt/C on carbon fiber paper electrodes kept at 0.70 V vs. RHE in $O_2$-saturated 1 M KOH. Catalyst loading is ~0.24 mg/cm$^2$ for all samples including Pt/C.

To investigate the ORR catalyst performance under conditions close to those present in a fuel cell setup, catalysts were loaded on Teflon-coated carbon fiber paper (CFP) (at a loading of ~0.24 mg/cm$^2$ for all samples including Pt/C) to measure their $iR$-compensated polarization curves[13]. The $MnCo_2O_4$/N-rmGO hybrid afforded significantly higher current densities than each component alone ($MnCo_2O_4$ or N-rmGO) or their physical mixture ($MnCo_2O_4$ + N-rmGO or $MnCo_2O_4$ + carbon black) (Figure 3a). In 1 M KOH at 0.7 V vs. RHE, the $MnCo_2O_4$/N-rmGO hybrid afforded an ORR current density of 151 mA/cm$^2$, which is about 7 times that of the mixture of $MnCo_2O_4$ + N-rmGO (22 mA/cm$^2$). In electrochemical impedance spectroscopy measurements, the $MnCo_2O_4$/N-rmGO hybrid showed a much smaller semicircle than the physical mixture and free $MnCo_2O_4$ nanoparticles within the ORR active potential range (Figure 3b & Fig. S6). This suggested a much smaller charge transfer resistance for the hybrid materials during ORR, due to the intimate nanocrystal-graphene coupling that enhanced electron transport between spinel oxide and graphene[19]. The $MnCo_2O_4$/N-rmGO hybrid showed higher current density than pure $Co_3O_4$/N-rmGO hybrid, consistent with CV and RDE data (Figure 2a & 2b). Impressively, the current density of $MnCo_2O_4$/N-rmGO hybrid was similar to that of Pt/C catalyst at low overpotentials and exceeded that of Pt/C at higher overpotential (Figure 3c). A similar trend was also observed in 0.1 M KOH (Fig. S7).

Besides a high activity, the $MnCo_2O_4$/N-rmGO hybrid also exhibited excellent stability as measured by chronoamperometric measurements. At a constant voltage of 0.70 V vs. RHE, the ORR current density produced in the hybrid catalyst decreased by only 3.5 % over 20,000 s of continuous operation, while the corresponding physical mixture sample and Pt/C catalysts exhibited ~ 25 % and 33 % decreases in current density, respectively (Figure 3d).

X-ray near edge structure (XANES) measurements using the total electron yield were employed to investigate the interaction between the spinel oxide nanoparticles and N-doped reduced graphene oxide and to examine the impact of Mn substitution in the hybrid materials. Absorption features at 285.5 eV (graphitic C-C π*) and at 291.7 and 292.7 eV (graphitic C-C σ*) in the carbon K-edge XANES (Figure 4a) correspond to graphene with crystalline structural order[20]. The transition at ~ 288 eV and the nearby shoulders before and after this peak are characteristic of defects in graphene associated with nitrogen and oxygen functional groups, respectively (Figure 4a). Presumably, the features at ~287.5 eV are due to C=N π*,[21] the features at ~288.5 eV are due to O-C=O π*,[20] and the features at ~289.5 eV are due to C-OH and C-O-C σ*.[22] The much enhanced carbon K-edge peak at ~288 eV and shoulder features near this peak in the hybrid materials (Figure 4a) suggested significantly perturbed bonding between carbon-oxygen and carbon-nitrogen,[23] likely resulted from bonding of the oxygen and nitrogen to metal atoms in the oxide nanoparticles. Such enhanced features had also been observed with metal oxides formed on oxidized carbon nanotubes[21a]. The nitrogen K-edge XANES (Figure 4b) of the hybrid exhibited enhanced π* peaks in the range of 398-402 eV (398.7 eV – pyridinic, 400 eV – pyrrolic or amino, 402 eV – graphitic)[24] and broadened σ* peak at ~407 eV compared to the nitrogen K-edge XANES of N-rmGO. Enhancement of the π* peak could be attributed to coordination of nitrogen to metal in the hybrid[21], causing charge transfer from nitrogen to metal. The broadening of σ* peak suggested stronger C-N bonding, which could be a result of covalent N-rmGO and metal oxide coupling[25]. All these suggested the formation of covalent interfacial metal-O-C and metal-N-C bonds in the hybrids. Such intimate interaction renders the oxide nanoparticles highly conducting and electrochemically active, much more so than the physical mixture of metal oxide nanoparticles and N-doped graphene.

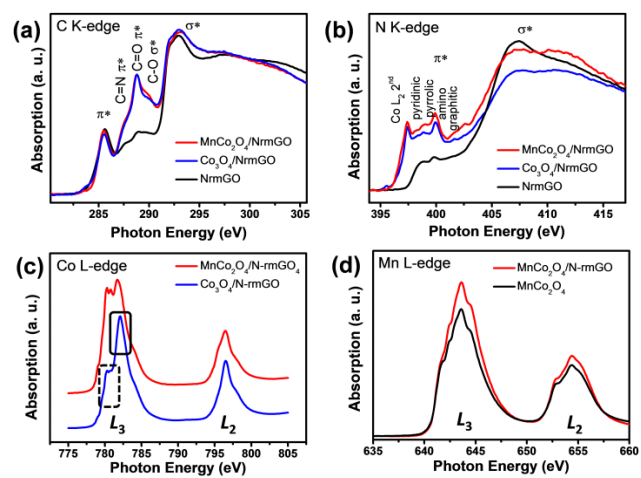

**Figure 4.** (a) C K-edge XANES of $MnCo_2O_4$/N-rmGO hybrid (red curve), $Co_3O_4$/N-rmGO hybrid (blue curve) and N-rmGO (black curve); (b) N K-edge XANES of $MnCo_2O_4$/N-rmGO hybrid (red curve), $Co_3O_4$/N-rmGO hybrid (blue curve) and N-rmGO (black curve). Peak at 397 eV was attributed to the second order photon excited Co $L_2$. (c) Co L-edge XANES of $MnCo_2O_4$/N-rmGO hybrid (red curve), and $Co_3O_4$/N-rmGO hybrid (blue curve). Spectra were shifted vertically to clearly show the difference in the Co $L_3$ fine features of the $Co_3O_4$ and $MnCo_2O_4$ hybrids. The solid and dashed black boxes mark the high and low energy states respectively in the Co $L_3$ region. (d) Mn L-edge XANES of $MnCo_2O_4$/N-rmGO hybrid (red curve) and $MnCo_2O_4$ (black curve).

L-edge XANES involves excitation of 2p electrons into unoccupied states of 3d character and were employed to probe the valence and the coordination environment of the Co and Mn. The Co $L_3$ edge (transition from Co $2p_{3/2}$ to Co 3d projected unoccupied states) of the pure $Co_3O_4$/N-rmGO hybrid (Figure 4c) was found to be consistent with a mixture of octahedrally coordinated $Co^{3+}$ and tetrahedrally coordinated $Co^{2+}$ in a 2:1 ratio[26]. The ratio of high energy/low energy peaks (marked by solid and dashed black boxes respectively in Figure 4c) decreased markedly for $MnCo_2O_4$/N-rmGO hybrid compared to $Co_3O_4$/N-rmGO, suggesting an increase in the ratio of $Co^{2+}/Co^{3+}$ in the Mn substituted hybrid[27]. A relative increase of $Co^{2+}$ ($3d^7$) over $Co^{3+}$ ($3d^6$) due to Mn substitution led to a relative decrease in unoccupied high-energy Co 3d states, thus lowering the high/low energy peak ratio in the Co $L_3$ edge region. The Mn L-edge XANES of $MnCo_2O_4$/N-rmGO hybrid was very close to that of a $Mn_2O_3$ reference (Figure 4d and Fig. S8), suggesting Mn was mainly in the 3+ state and has substituted for the $Co^{3+}$ in the oxide, which led to a spinel structure of $Co^{II}(Co^{III}Mn^{III})O_4$. A slight shoulder observed at 644.5 eV in Mn L edge may suggest the existence of small amount of $Mn^{4+}$.[28] Note that $Mn^{3+}$ and $Mn^{4+}$ species were considered to be active sites for ORR[16] and were more active than Co species[10]. Although both $MnCo_2O_4$/N-rmGO and $Co_3O_4$/N-rmGO hybrids showed similar BET surface area (210 m$^2$/g for $Co_3O_4$ and 208 m$^2$/g for $MnCo_2O_4$), $MnCo_2O_4$/N-rmGO showed a larger loop compared to $Co_3O_4$ hybrid in the CV data measured under $N_2$ (Figure 2a), suggesting Mn doping increased the electro-chemically active surface area of the hybrid. The increased activity of the active sites and higher electrochemically active surface area were likely factors in the enhanced catalytic activity in the Mn substituted hybrid.

$MnCo_2O_4$/N-rmGO hybrid with larger nanoparticle size of ~10 nm was prepared and the material showed lower ORR activity (Figure S9a) than the 5 nm ones, indicating smaller $MnCo_2O_4$ particle size was more desirable for higher electrochemical activity. We also examined the ORR activities of $MnCo_2O_4$/N-rmGO hybrids with various $MnCo_2O_4$ contents (Figure S9b). An optimum range of $MnCo_2O_4$ content between 65% - 80 wt% was found to afford similarly high ORR performance from carbon fiber paper measurements. Out of this range, too low a $MnCo_2O_4$ content could lead to fewer ORR active sites in the hybrids, while too high a $MnCo_2O_4$ content could result in aggregation of nanoparticles and even free growth, which were less active than nanoparticles directly grown on graphene sheets.

We synthesized hybrids with various Mn/Co ratios by varying the reactant ratio of cobalt acetate/manganese acetate in the nucleation step of our synthesis and characterized their ORR catalytic performance (Fig. S10). The $MnCo_2O_4$/N-rmGO hybrid showed higher ORR current density than $Mn_{0.6}Co_{2.4}O_4$/N-rmGO. However, increasing the Mn/Co ratio above 1/2 led to decrease of ORR current density with the $Mn_{1.5}Co_{1.5}O_4$/N-rmGO, $Mn_2CoO_4$/N-rmGO and $Mn_3O_4$/N-rmGO hybrids. Nanoparticles sizes of ~ 10-20 nm formed on N-rmGO sheets were observed in $Mn_{1.5}Co_{1.5}O_4$/N-rmGO and $Mn_2CoO_4$/N-rmGO hybrids, and the size further increased with higher Mn/Co ratios (Fig. S11). The increase of nanoparticle size in the hybrid with higher Mn/Co ratio was likely due to a reduced coordination effect of ammonia to Mn cations than to Co cations during the nucleation step. XRD revealed that the oxide in $Mn_2CoO_4$/N-rmGO was in a tetragonal spinel phase (Fig. S12) rather than cubic spinel, consistent with previous reports of spinel manganese-cobalt oxide[14]. It had been reported that the tetragonal spinel exhibited lower intrinsic ORR activity than the cubic phase[15]. This and the increase of particle size could explain decreased ORR catalytic activity in Mn substituted hybrid with Mn/Co ratio > 0.5. This further suggested that the active site in the hybrid materials was spinel oxide formed on graphene sheets, and the phase and size of nanoparticles were important for optimum ORR performance of the hybrid.

Lastly, we investigated electrocatalysis of the oxygen evolution reaction (OER) with the catalysts loaded on carbon fiber paper (Fig. 5). With Mn substitution, $MnCo_2O_4$/N-rmGO hybrid exhibited lower OER currents than the pure $Co_3O_4$/N-rmGO hybrid. This was consistent with previous XANES observations that $Co^{3+}$ was considered to be the active sites for OER[9], which was reduced due to $Mn^{3+}$ substitution. Still, the $MnCo_2O_4$/N-rmGO hybrid showed much higher OER catalytic activity than the physical mixture, suggesting that the $MnCo_2O_4$ hybrid is an efficient bi-functional catalyst for ORR and OER.

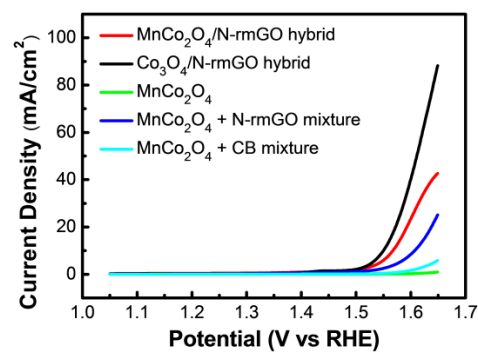

**Figure 5.** Oxygen evolution currents of $MnCo_2O_4$/N-rmGO hybrid, $Co_3O_4$/N-rmGO hybrid, $MnCo_2O_4$, $MnCo_2O_4$ + N-rmGO mixture and $MnCo_2O_4$ + CB mixture dispersed on carbon fiber paper in 1 M KOH electrolyte (catalyst loading ~ 0.24 mg/cm$^2$ for all samples).

## CONCLUSION

In conclusion, by combining nanoparticle nucleation and growth on graphene oxide sheets and cation substitution of spinel metal oxide nanoparticles, we have developed a $MnCo_2O_4$/N-doped graphene hybrid material for highly efficient ORR electrocatalysis in alkaline conditions. The nucleation and growth method results in covalent coupling between spinel oxide nanoparticles and N-doped reduced graphene oxide sheets, affording much higher activity and stronger durability than the physical mixture of nanoparticles and N-rmGO. Mn substitution increased the activity of catalytic sites of the hybrid materials, further boosting the ORR activity compared to the pure $Co_3O_4$/N-rmGO hybrid. At the same mass loading, the $MnCo_2O_4$/N-rmGO hybrid can outperform Pt/C in ORR current density at potential < 0.75 V vs. RHE with superior stability to Pt/C. Mn substitution is also found to mediate the nanoparticle size and phase, affecting the hybrid catalytic performance. Rational design of surface structure and conductivity through cationic substitution and covalent coupling with carbon substrates should benefit the design of advanced ORR electrocatalyst for energy conversion and storage.

## EXPERIMENTAL SECTION

**Synthesis of mildly oxidized graphene oxide (mGO).** mGO was made by a modified Hummers method using a lower concentration of oxidizing agent[12c, 29]. Graphite flakes (1 g, Superior Graphite Co.) were grounded with NaCl (20 g) for 10-15 minutes. Afterwards, the NaCl was washed away by repeatedly rinsing with water in a vacuum filtration apparatus. The remaining graphite was dried in an oven at 70°C for 30 minutes. The dried solid was transferred to a 250 ml round bottom flask. 23 ml of concentrated sulfuric acid was added and the mixture was stirred at room temperature for 24 hours. Next, the flask was heated in an oil bath at 40°C. 100 mg of NaNO$_3$ was added to the suspension and allowed to dissolve for 5 minutes. This step was followed by the slow addition of 500 mg of KMnO$_4$ (as opposed to 3 g for Hummers' GO), keeping the reaction temperature below 45°C. The solution was allowed to stir for 30 minutes. Afterwards, 3 ml of water was added to the flask, followed by another 3 ml after 5 minutes. After another 5 minutes, 40 ml of water was added. 15 minutes later, the flask was removed from the oil bath and 140 ml of water and 10 ml of 30% $H_2O_2$ were added to end the reaction. This suspension was stirred at room temperature for 5 minutes. It was then repeatedly centrifuged and washed with 5% HCl solution twice, followed by rinsing with copious amounts of water. The final precipitate was dispersed in 100 ml of water and bath sonicated for 30 min. Any indispensable solid was crushed down by a centrifugation at 5000 rpm for 5 minutes, and a brown homogeneous supernatant was collected.

**Synthesis of $MnCo_2O_4$/N-rmGO hybrid, N-rmGO, $MnCo_2O_4$ nanoparticles and $Mn_xCo_{3-x}O_4$/N-rmGO hybrids.** mGO was collected from the aqueous solution by centrifugation and redispersed in anhydrous ethanol (EtOH). The concentration of the final mGO EtOH suspension was ~0.33 mg/ml (concentration of our mGO stock suspension was determined by measuring the mass of the mGO lyophilized from a certain volume of the suspension). For the first step in the synthesis of the hybrid, 0.533 ml of 0.6 M $Co(OAc)_2$ aqueous solution and 0.266 ml of 0.6 M $Mn(OAc)_2$ aqueous solution was added to 24 ml of mGO EtOH suspension, followed by the addition of 0.3 ml of water and 0.5 mL of $NH_4OH$ at RT. The reaction was kept at 80 °C with stirring for 20 h. After that, the reaction mixture from the first step was transferred to a 40 mL autoclave for hydrothermal reaction at 150°C for 3 h. This hydrothermal step also reduced mGO to rmGO. The resulting product was collected by centrifugation and washed with ethanol and water. The $MnCo_2O_4$/N-rmGO hybrid product was ~36 mg after lyophilization.

N-rmGO was made through the same steps as $MnCo_2O_4$/N-rmGO without adding any Co & Mn salt in the first step. This produced N-doped reduced GO due to the hydrothermal step. Free $MnCo_2O_4$ nanoparticles were made through the same steps as making $MnCo_2O_4$/N-rmGO without adding any mGO in the first step.

$Mn_xCo_{3-x}O_4$/N-rmGO hybrids were synthesized by the same method as $MnCo_2O_4$/N-rmGO hybrid by adjusting $Co(OAc)_2$/ $Mn(OAc)_2$ reactants ratio in the first step.

**Sample preparation for SEM, TEM, XRD & surface area measurements.** SEM samples were prepared by drop-drying the samples from their aqueous suspensions onto silicon substrates. TEM samples were prepared by drop-drying the samples from their diluted aqueous suspensions onto copper grids. XRD samples were prepared by drop-drying the samples from their aqueous suspensions onto glass substrates. Surface area of the samples was measured by Brunauer-Emmett-Teller (BET) nitrogen adsorption-desorption isotherms at 77 K with Micromeritics ASAP 2020 instrument.

**Electrochemical measurements**

1. **Cyclic voltammetry (CV).** 4 mg of catalyst and 13-87 µL of 5 wt% Nafion solution (13 µL for hybrids, 15 wt% of Nafion to catalyst ratio; 87 µL for N-rmGO, 100 wt% of Nafion to catalyst ratio) were dispersed in 1 ml of 2.5:1 v/v water/isopropanol mixed solvent by at least 30 min sonication to form a homogeneous ink. Pt/C ink was prepared by dispersing 4 mg Pt/C (20 wt% Pt on Vulcan XC-72) in 1 mL EtOH with 35 µl of 5 wt% Nafion solution (40 wt% of Nafion to catalyst ratio) by at least 30 min sonication. Then 5 µL of the catalyst ink (containing 20 µg of catalyst) was loaded onto a glassy carbon electrode of 5 mm in diameter (loading ~ 0.10 mg/$cm^2$). Cyclic voltammetry was conducted with a CHI 760 D potentiostat in a three electrode electrochemical cell using saturated calomel electrode as the reference electrode, a graphite rod as the counter electrode and the sample modified glassy carbon electrode as the working electrode. Electrolyte was saturated with oxygen by bubbling $O_2$ prior to the start of each experiment. A flow of $O_2$ was maintained over the electrolyte during the recording of CVs in order to ensure its continued $O_2$ saturation. The working electrode was cycled at least 10 times before data were recorded at a scan rate of 5 mVs$^{-1}$. In control experiments, CV measurements were also performed in $N_2$ by switching to $N_2$ flow through the electrochemical cell.

2. **Rotating disk electrode (RDE) measurement.** The catalyst modified working electrode was prepared by the same method as CV. The working electrode was scanned cathodically at a rate of 5 mVs$^{-1}$ with varying rotating speed from 400 rpm to 2025 rpm. Koutecky–Levich plots ($J^{-1}$ vs. $\omega^{-1/2}$) were analyzed at various electrode potentials. The slopes of their best linear fit lines were used to calculate the number of electrons transferred (n) on the basis of the Koutecky-Levich equation[18]:

$$\frac{1}{J} = \frac{1}{J_L} + \frac{1}{J_K} = \frac{1}{B\omega^{1/2}} + \frac{1}{J_K}$$

$$B = 0.62nFC_oD_o^{2/3}v^{-1/6}; \quad J_K = nFkC_o$$

where $J$ is the measured current density, $J_K$ and $J_L$ are the kinetic- and diffusion- limiting current densities, $\omega$ is the angular velocity, $n$ is transferred electron number, $F$ is the Faraday constant, $C_o$ is the bulk concentration of $O_2$, $v$ is the kinematic viscosity of the electrolyte, and $k$ is the electron-transfer rate constant. For the Tafel plot, the kinetic current was calculated from the mass-transport correction of RDE by:

$$J_K = \frac{J \times J_L}{(J_L - J)}$$

3. **Rotating ring-disk electrode (RRDE) measurement.** For the RRDE measurements, catalyst inks and electrodes were prepared by the same method as RDE's. The ink was dried slowly in air and the drying condition was adjusted by trial and error until a uniform catalyst distribution across the electrode surface was obtained. The disk electrode was scanned cathodically at a rate of 5 mVs$^{-1}$ and the ring potential was constant at 1.3 V vs. RHE. The % $HO_2^-$ and the electron transfer number (n) were determined by the followed equations[17]:

$$HO_2^- = 200 \times \frac{I_r/N}{I_d + I_r/N}$$

$$n = 4 \times \frac{I_d}{I_d + I_r/N}$$

where $I_d$ is disk current, $I_r$ is ring current and N is current collection efficiency (N) of the Pt ring. N was determined to be 0.40 from the reduction of $K_3Fe[CN]_6$.

4. **Oxygen electrode activities on carbon fiber paper.** For measurements on carbon fiber paper, the working electrode was prepared by loading ~ 0.24 mg of catalyst (for hybrid catalysts and Pt/C) on 1 $cm^2$ carbon fiber paper (purchased from Fuel Cell Store) from its 1 mg/ml ethanol dispersion with a 1:10 Nafion-to-catalyst ratio. It was cycled at least 20 times between 0 and 0.6 V vs. SCE before data were recorded at a scan rate of 5mVs$^{-1}$ for ORR measurement. To obtain OER activities in 1 M KOH, the working electrode was scanned from 0 V to 0.6 V vs. SCE. All the data from carbon fiber paper were iR-compensated.

**5. RHE calibration** We used saturated calomel electrode (SCE) as the reference electrode in all measurements. It was calibrated with respect to reversible hydrogen electrode (RHE). The calibration was performed in the high purity hydrogen saturated electrolyte with a Pt wire as the working electrode. CVs were run at a scan rate of 1 mV s$^{-1}$, and the average of the two potentials at which the current crossed zero was taken to be the thermodynamic potential for the hydrogen electrode reactions. In 1 M KOH, E (RHE) = E (SCE) + 1.051 V.

**XANES measurements**. XANES measurements were performed at the SGM beamline of the Canadian Light Source. Powder samples were held by indium foil. XANES were recorded in the surface sensitive total electron yield (TEY) with use of specimen current. Data were first normalized to the incident photon flux $I_0$ measured with a refreshed gold mesh at SGM prior to the measurement. After background correction, the XANES are then normalized to the edge jump, the difference in absorption coefficient just below and at a flat region above the edge (300, 415, 805, and 660 eV for C, N, Co and Mn respectively).

## AUTHOR INFORMATION

### Corresponding Author

* hdai@stanford.edu

### Author Contributions

§These authors contributed equally.

## ACKNOWLEDGMENT

This work was supported by a Stinehart Grant for Energy Research at Stanford from the Stanford Precourt Institute for Energy. CLS is supported by the NSERC, NRC, CIHR of Canada, the Province of Saskatchewan, WEDC, and the University of Saskatchewan.

## REFERENCES

1. Gewirth, A. A.; Thorum, M. S., *Inorg. Chem.* **2010**, *49*, 3557-3566.
2. Armand, M.; Tarascon, J. M., *Nature* **2008**, *451*, 652-657.
3. (a) Stamenkovic, V.; Mun, B. S.; Mayrhofer, K. J. J.; Ross, P. N.; Markovic, N. M.; Rossmeisl, J.; Greeley, J.; Norskov, J. K., *Angew. Chem. Int. Ed.* **2006**, *45*, 2897-2901; (b) Srivastava, R.; Mani, P.; Hahn, N.; Strasser, P., *Angew. Chem. Int. Ed.* **2007**, *46*, 8988-8991; (c) Morozan, A.; Jousselme, B.; Palacin, S., *Energy & Environmental Science* **2011**, *4*, 1238-1254.
4. Spendelow, J. S.; Wieckowski, A., *Phys. Chem. Chem. Phys.* **2007**, *9*, 2654-2675.
5. (a) Meadowcr.Db, *Nature* **1970**, *226*, 847-&; (b) Sugawara, M.; Ohno, M.; Matsuki, K., *J. Mater. Chem.* **1997**, *7*, 833-836; (c) Verma, A.; Jha, A. K.; Basu, S., *J. Power Sources* **2005**, *141*, 30-34.
6. (a) Piana, M., Catanorchi, S. & Gasteiger, H. A., *Electrochem. Soc. Trans.* **2008**, *16*, 2045-2055; (b) Meng, H.; Jaouen, F.; Proietti, E.; Lefevre, M.; Dodelet, J. P., *Electrochem. Commun.* **2009**, *11*, 1986-1989.
7. Su, D. S.; Zhang, J.; Frank, B.; Thomas, A.; Wang, X.; Paraknowitsch, J.; Schloegl, R., *ChemSusChem* **2010**, *3*, 169-180.
8. Neburchilov, V.; Wang, H.; Martin, J. J.; Qu, W., *J. Power Sources* **2010**, *195*, 1271-1291.
9. Hamdani, M.; Singh, R. N.; Chartier, P., *Int. J. Electrochem. Sci.* **2010**, *5*, 556-577.
10. De Koninck, M.; Marsan, B., *Electrochim. Acta* **2008**, *53*, 7012-7021.
11. Bidault, F.; Brett, D. J. L.; Middleton, P. H.; Brandon, N. P., *J. Power Sources* **2009**, *187*, 39-48.
12. (a) Liang, Y.; Wang, H.; Casalongue, H. S.; Chen, Z.; Dai, H., *Nano Res.* **2010**, *3*, 701-705; (b) Wang, H.; Casalongue, H. S.; Liang, Y.; Dai, H., *J. Am. Chem. Soc.* **2010**, *132*, 7472-7477; (c) Wang, H.; Cui, L.-F.; Yang, Y.; Casalongue, H. S.; Robinson, J. T.; Liang, Y.; Cui, Y.; Dai, H., *J. Am. Chem. Soc.* **2010**, *132*, 13978-13980; (d) Li, Y.; Wang, H.; Xie, L.; Liang, Y.; Hong, G.; Dai, H., *J. Am. Chem. Soc.* **2011**, *133*, 7296-7299; (e) Wang, H.; Yang, Y.; Liang, Y.; Cui, L.-F.; Casalongue, H. S.; Li, Y.; Hong, G.; Cui, Y.; Dai, H., *Angew. Chem. Int. Ed.* **2011**, *50*, 7364-7368; (f) Wang, H.; Liang, Y.; Mirfakhrai, T.; Chen, Z.; Casalongue, H. S.; Dai, H., *Nano Res.* **2011**, *4*, 729-736.
13. Liang, Y.; Li, Y.; Wang, H.; Zhou, J.; Wang, J.; Regier, T.; Dai, H., *Nat. Mater.* **2011**, *10*, 780-786.
14. Devidales, J. L. M.; Vila, E.; Rojas, R. M.; Garciamartinez, O., *Chem. Mater.* **1995**, *7*, 1716-1721.
15. Cheng, F.; Shen, J.; Peng, B.; Pan, Y.; Tao, Z.; Chen, J., *Nature Chem.* **2011**, *3*, 79-84.
16. Rios, E.; Gautier, J. L.; Poillerat, G.; Chartier, P., *Electrochim. Acta* **1998**, *44*, 1491-1497.
17. Paulus, U. A.; Schmidt, T. J.; Gasteiger, H. A.; Behm, R. J., *J. Electroanal. Chem.* **2001**, *495*, 134-145.
18. Bard, A. J.; Faulkner, L. R., *Electrochemical Methods: Fundamentals and Aplications*. Wiley: New York, 2001.
19. Genies, L.; Bultel, Y.; Faure, R.; Durand, R., *Electrochim. Acta* **2003**, *48*, 3879-3890.
20. Zhou, J. G.; Wang, J.; Sun, C. L.; Maley, J. M.; Sammynaiken, R.; Sham, T. K.; Pong, W. F., *J. Mater. Chem.* **2011**, *21*, 14622-14630.
21. Kelly, D. N.; Schwartz, C. P.; Uejio, J. S.; Duffin, A. M.; England, A. H.; Saykally, R. J., *J. Chem. Phys.* **2010**, *133*.
22. Saxena, S.; Tyson, T. A.; Negusset, E., *Journal of Physical Chemistry Letters* **2010**, *1*, 3433-3437.
23. (a) Zhou, J.; Wang, J.; Fang, H.; Sham, T.-K., *J. Mater. Chem.* **2011**, *21*, 5944-5949; (b) Zhang, L.-S.; Liang, X.-Q.; Song, W.-G.; Wu, Z.-Y., *Phys. Chem. Chem. Phys.* **2010**, *12*, 12055-12059.
24. Leinweber, P.; Kruse, J.; Walley, F. L.; Gillespie, A.; Eckhardt, K.-U.; Blyth, R. I. R.; Regier, T., *J. Synchrotron Rad.* **2007**, *14*, 500-511.
25. Zhou, J.; Zhou, X.; Li, R.; Sun, X.; Ding, Z.; Cutler, J.; Sham, T.-K., *Chem. Phys. Lett.* **2009**, *474*, 320-324.
26. Morales, F.; de Groot, F. M. F.; Glatzel, P.; Kleimenov, E.; Bluhm, H.; Havecker, M.; Knop-Gericke, A.; Weckhuysen, B. M., *J. Phys. Chem. B* **2004**, *108*, 16201-16207.
27. Zheng, F.; Alayoglu, S.; Guo, J.; Pushkarev, V.; Li, Y.; Glans, P.-A.; Chen, J.-l.; Somorjai, G., *Nano Lett.* **2011**, *11*, 847-853.
28. Grush, M. M.; Chen, J.; Stemmler, T. L.; George, S. J.; Ralston, C. Y.; Stibrany, R. T.; Gelasco, A.; Christou, G.; Gorun, S. M.; PennerHahn, J. E.; Cramer, S. P., *J. Am. Chem. Soc.* **1996**, *118*, 65-69.
29. Sun, X.; Liu, Z.; Welsher, K.; Robinson, J. T.; Goodwin, A.; Zaric, S.; Dai, H., *Nano Res.* **2008**, *1*, 203-212.

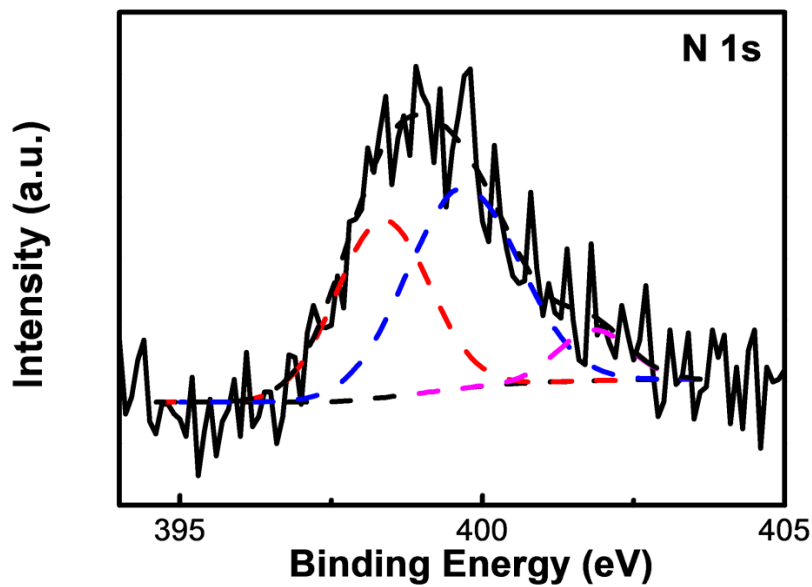

**Fig. S1**. A high resolution N1s XPS spectrum of $MnCo_2O_4$/N-rmGO hybrid. The peak was deconvoluted into pyridinic (red), pyrrolic/amino (blue), and graphitic (magenta) N peaks.

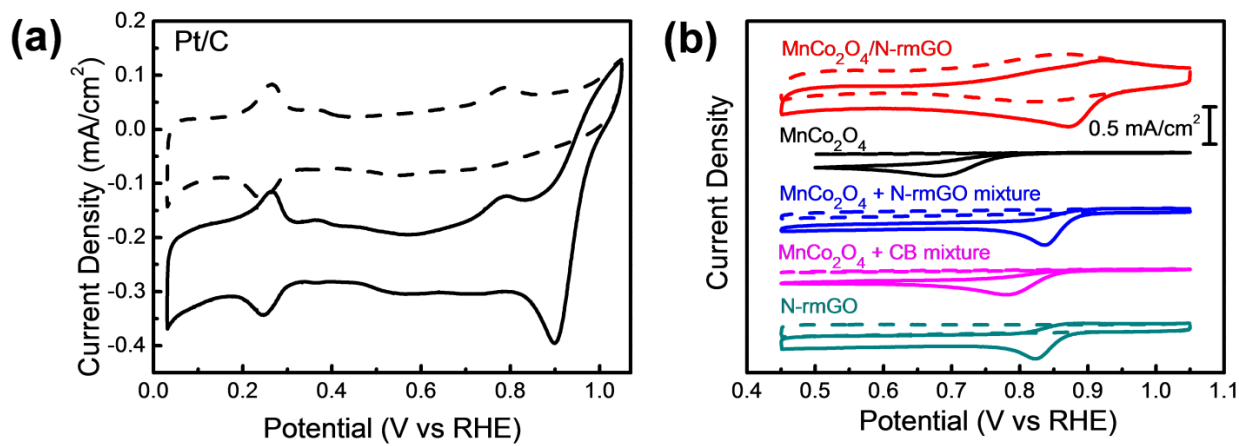

**Fig. S2**. (a) CV curves of Pt/C on glassy carbon electrodes in $O_2$-saturated (solid line) or $N_2$-saturated (dash line) 1 M KOH. (b) CV curves of $MnCo_2O_4$/N-rmGO hybrid, $MnCo_2O_4$, $MnCo_2O_4$+N-rmGO mixture, $MnCo_2O_4$+CB mixture and N-rmGO on glassy carbon electrodes in $O_2$-saturated (solid line) or $N_2$-saturated (dash line) 1 M KOH.

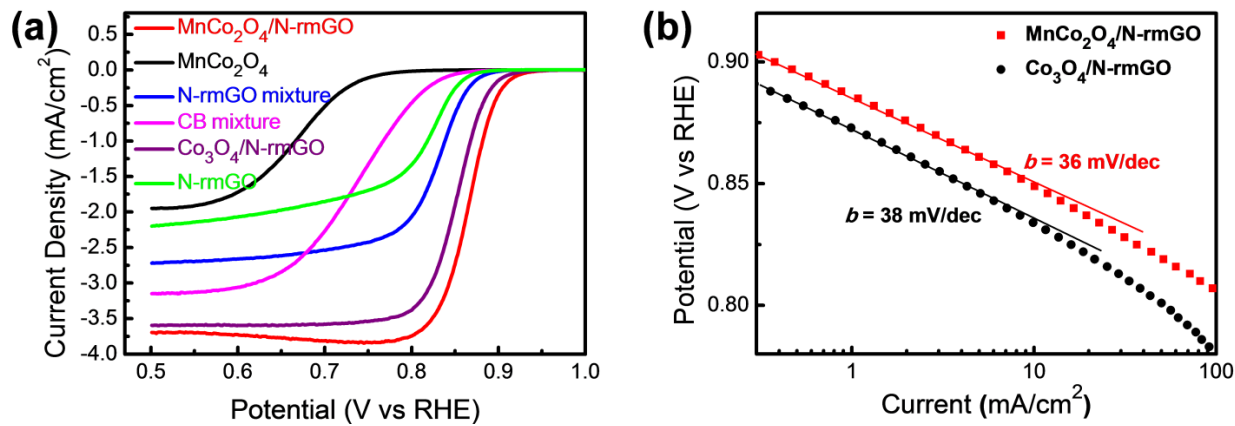

**Fig. S3**. (a) Rotating-disk electrode voltammograms of $MnCo_2O_4$/N-rmGO hybrid, $MnCo_2O_4$, $MnCo_2O_4$+N-rmGO mixture, $MnCo_2O_4$+CB mixture and N-rmGO in $O_2$-saturated 1 M KOH with a sweep rate of 5 mV/s at 1600 rpm (Loading is 0.1 mg/cm$^2$). (b) Tafel plots of $MnCo_2O_4$/N-rmGO and $Co_3O_4$/N-rmGO hybrids derived by the mass-transport correction of corresponding RDE data in (a).

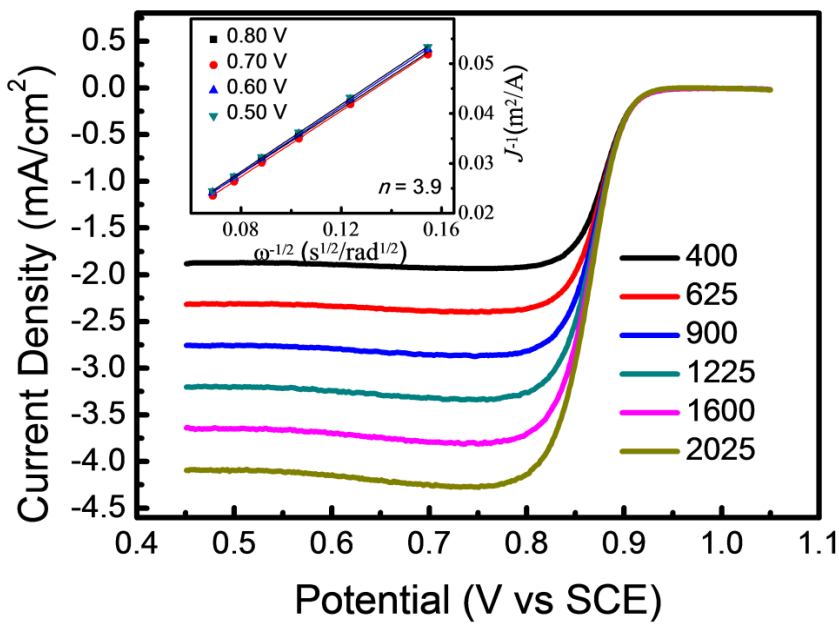

**Fig. S4**. Rotating-disk voltammograms of $MnCo_2O_4$/N-rmGO hybrid in $O_2$-saturated 1 M KOH with a sweep rate of 5 mV/s at different rotation rates indicated. The inset shows the corresponding Koutecky–Levich plots ($J^{-1}$ vs. $\omega^{-0.5}$) at different potentials.

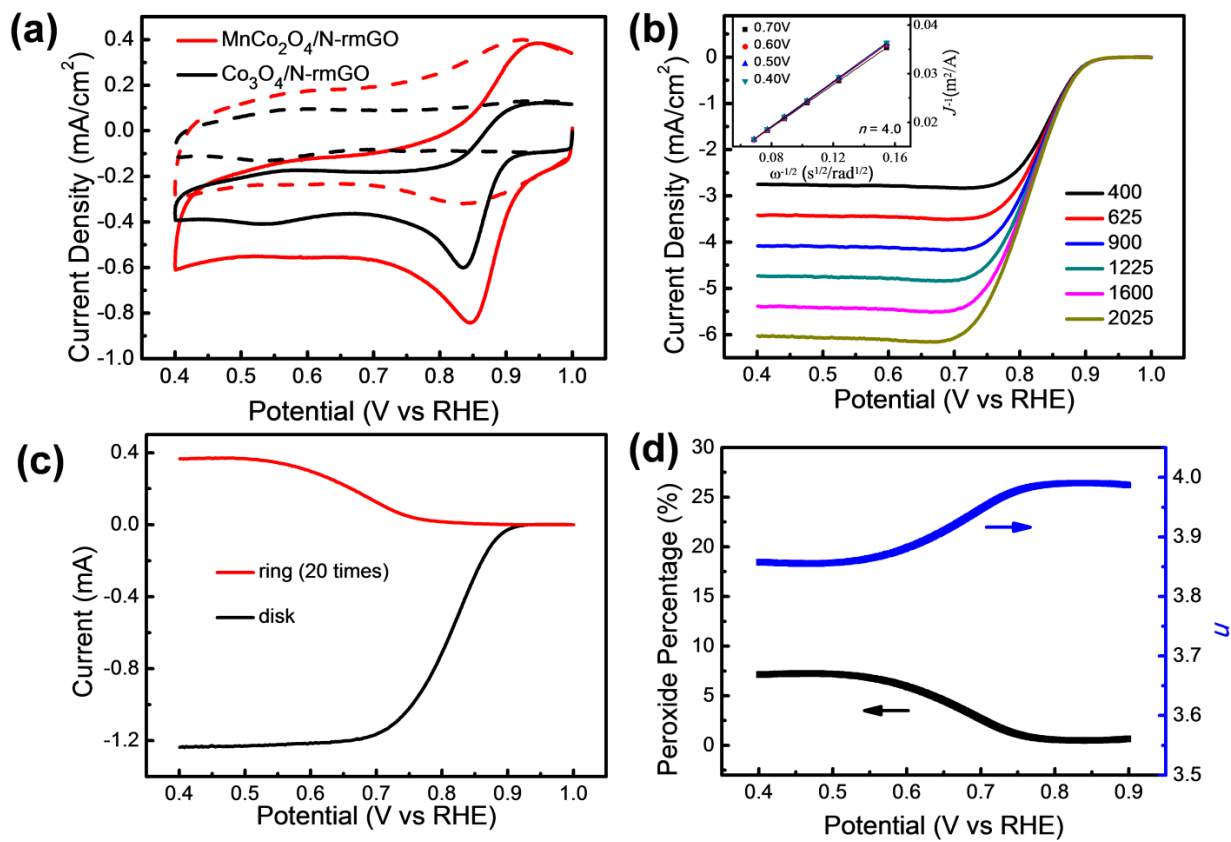

**Fig. S5**. (a) CV curves of MnCo$_2$O$_4$/N-rmGO hybrid & Co$_3$O$_4$/N-rmGO hybrid on glassy carbon electrodes in O$_2$-saturated (solid line) or N$_2$-saturated (dash line) 0.1 M KOH. (b) Rotating-disk voltammograms of MnCo$_2$O$_4$/N-rmGO hybrid in O$_2$-saturated 0.1 M KOH with a sweep rate of 5 mV/s at different rotation rates indicated. The inset shows the corresponding Koutecky–Levich plots ($J^{-1}$ vs. $\omega^{-0.5}$) at different potentials. (c). Rotating ring-disk electrode voltammogram of MnCo$_2$O$_4$/N-rmGO hybrid and MnCo$_2$O$_4$+N-rmGO mixture in O$_2$-saturated 0.1 M KOH at 1600 rpm. The disk potential was scanned at 5 mV/s and the ring potential was constant at 1.3 V vs RHE. (d) Percentage of peroxide (bottom) and the electron transfer number ($n$) (top) of MnCo$_2$O$_4$/N-rmGO hybrid and MnCo$_2$O$_4$+N-rmGO mixture at various potentials based on the corresponding RRDE data in (c). Catalyst loading was 0.10 mg/cm$^2$ for both samples.

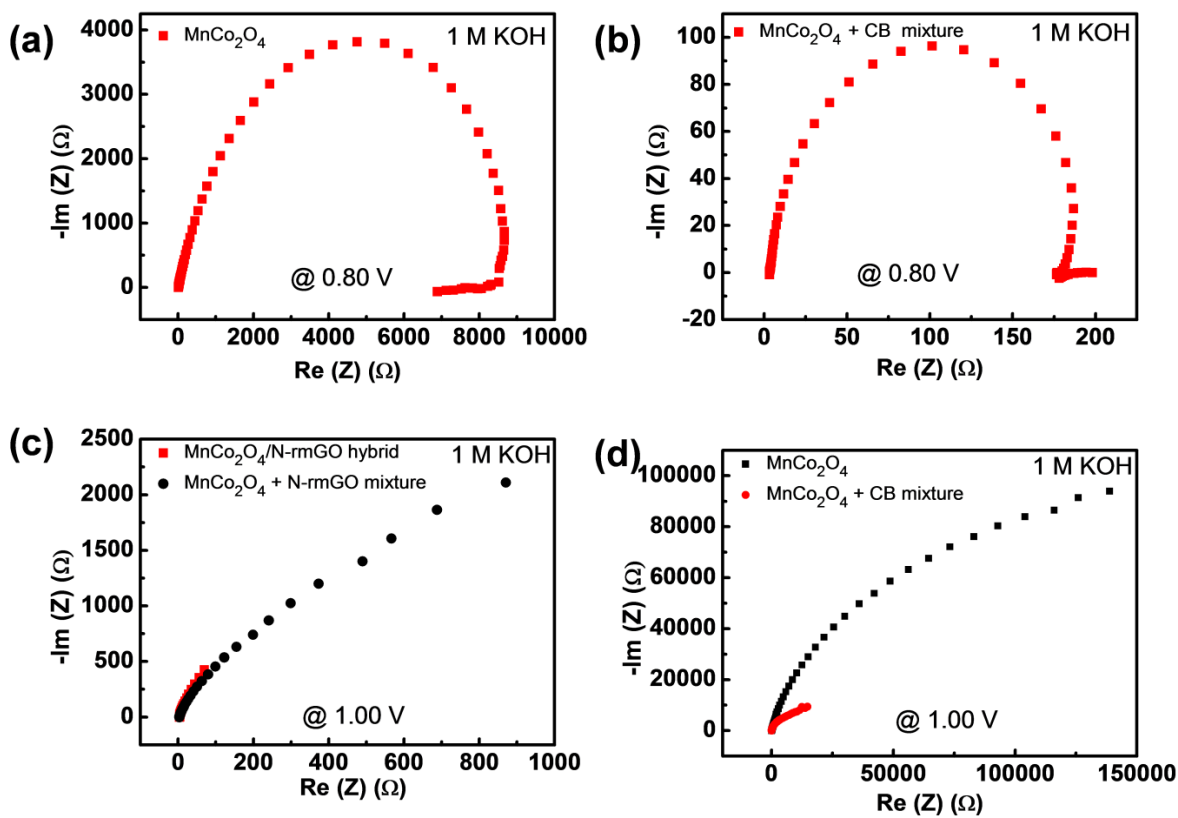

**Fig. S6.** Impedance study of hybrid and mixture. $MnCo_2O_4$ (a) and the mixture of $MnCo_2O_4$ and CB (b) at 0.8 V vs RHE. $MnCo_2O_4$/N-rmGO hybrid and the mixture of $MnCo_2O_4$ and N-rmGO (c) and $MnCo_2O_4$ and the mixture of $MnCo_2O_4$ and CB (d) at 1.0 V vs RHE.

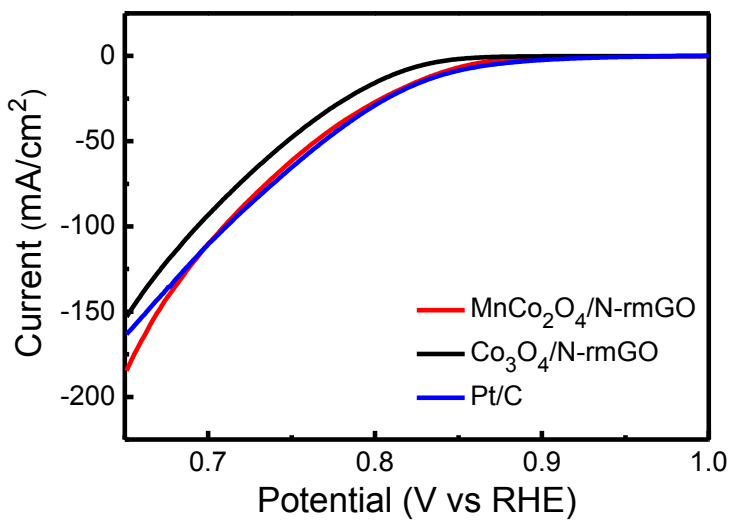

**Fig. S7.** Oxygen reduction polarization curves of $MnCo_2O_4$/N-rmGO hybrid, $Co_3O_4$/N-rmGO hybrid, and Pt/C on carbon fiber paper in 0.1 M KOH electrolyte. Catalyst loading is ~0.24 mg/cm$^2$ for all samples.

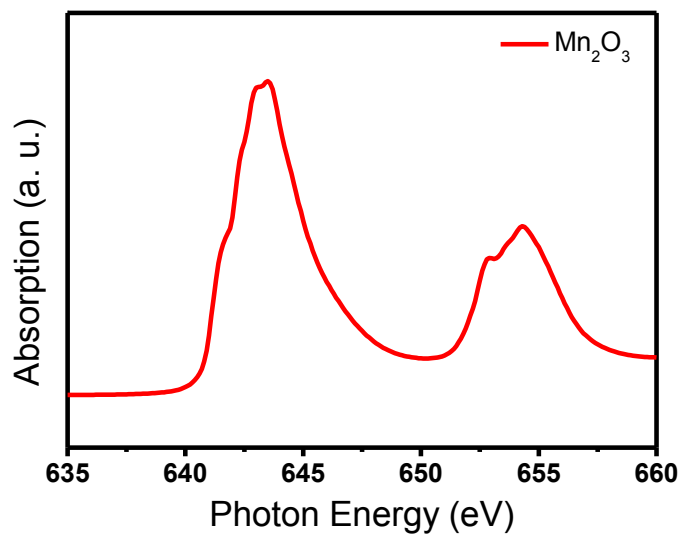

**Fig. S8.** Mn L-edge XANES of reference $Mn_2O_3$.

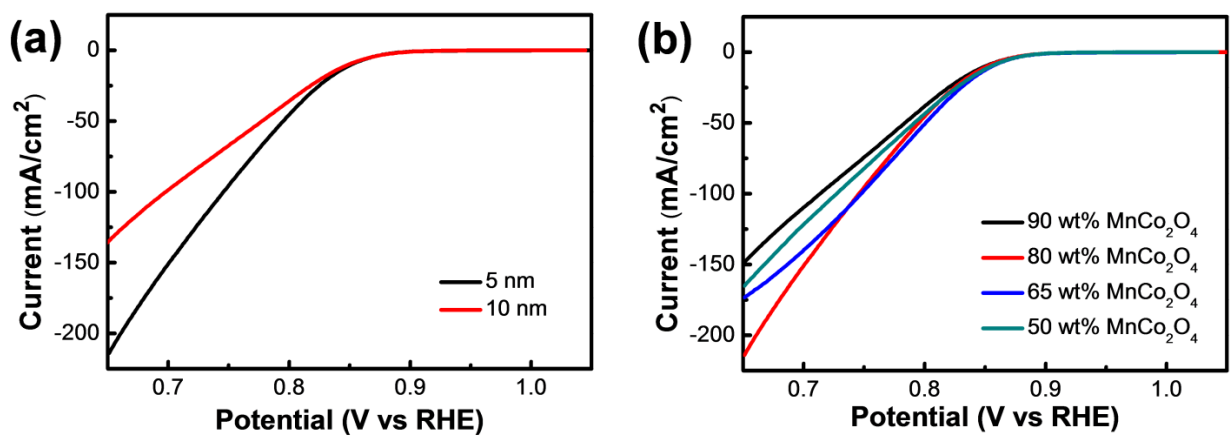

**Fig. S9.** (a) Oxygen reduction currents of $MnCo_2O_4$/N-rmGO hybrids with different $MnCo_2O_4$ particle size dispersed on carbon fiber paper in $O_2$-saturated 1 M KOH. The hybrid with 10 nm particle size was prepared by the similar method in experimental section, but with water/ethanol = 1/4 volume ratio in second step. Smaller $MnCo_2O_4$ particle size showed higher ORR activity. (b) Oxygen reduction currents of $MnCo_2O_4$/N-rmGO hybrids with various $MnCo_2O_4$ contents dispersed on carbon fiber paper in $O_2$-saturated 1 M KOH. The sample loading was 0.24 mg/cm$^2$. An optimum range of $MnCo_2O_4$ content between 65% - 80 wt% was found. Out of this range, too low $MnCo_2O_4$ content would lead to fewer ORR active sites in the hybrids, while too high $MnCo_2O_4$ content would result in aggregation of nanoparticles and even free growth, which are less active than nanoparticle in direct contact with graphene sheets.

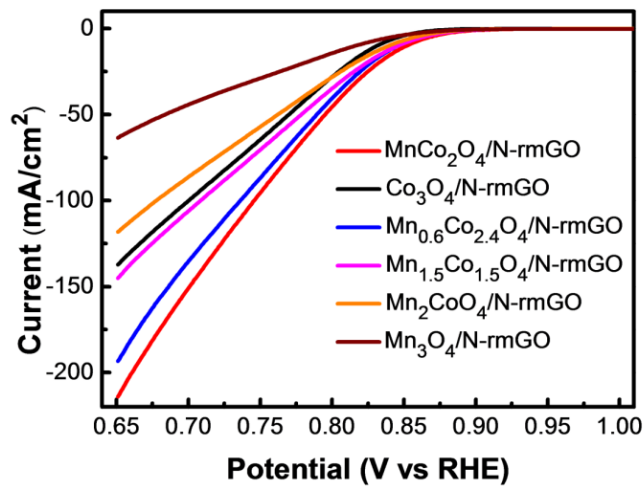

**Fig. S10.** Oxygen reduction polarization curves of MnCo$_2$O$_4$/N-rmGO hybrid, Co$_3$O$_4$/N-rmGO hybrid, Mn$_{0.6}$Co$_{2.4}$O$_4$/N-rmGO, Mn$_{1.5}$Co$_{1.5}$O$_4$/N-rmGO, Mn$_2$CoO$_4$/N-rmGO and Mn$_3$O$_4$/N-rmGO on carbon fiber paper in 1 M KOH electrolyte. Catalyst loading is ~0.24 mg/cm$^2$ for all samples.

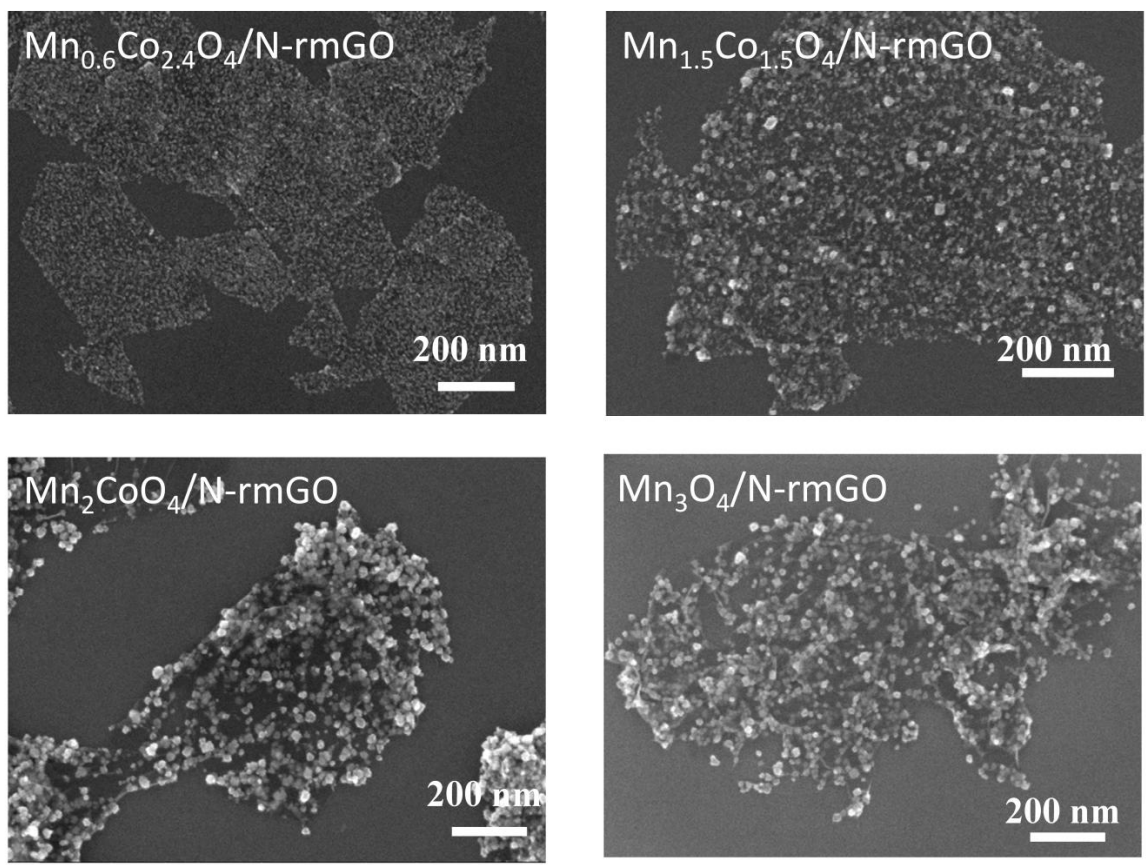

**Fig. S11**. Low magnification SEM images of hybrid with different Co/Mn ratio. (a) Mn$_{0.6}$Co$_{2.4}$O$_4$/N-rmGO; (b) Mn$_{1.5}$Co$_{1.5}$O$_4$/N-rmGO; (c) Mn$_2$CoO$_4$/N-rmGO and (d) Mn$_3$O$_4$/N-rmGO.

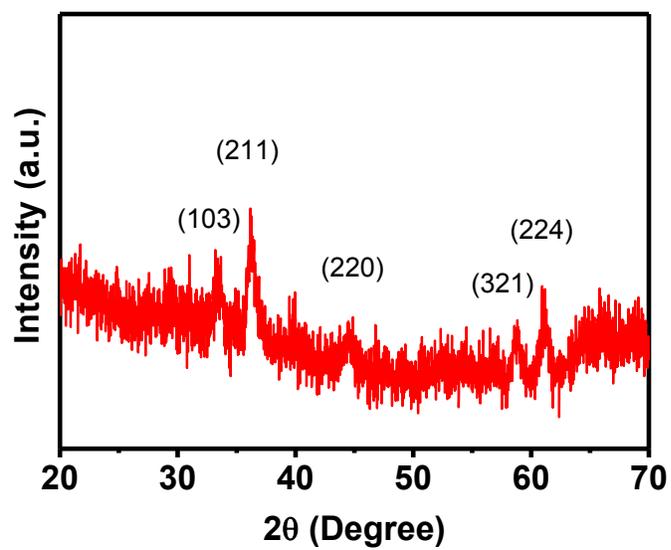

**Fig. S12**. XRD spectrum of a compacted film of $Mn_2CoO_4$/N-rmGO hybrid.